\documentclass{bmcart}

\usepackage{graphicx}
\usepackage{amsthm,amsmath}
\RequirePackage{hyperref}
\usepackage[utf8]{inputenc} 

\startlocaldefs
\endlocaldefs

\begin{document}

\begin{frontmatter}

\begin{fmbox}
\dochead{Research}


\title{Quality factor of a transmission line coupled coplanar waveguide resonator}


\author[
   addressref={MISIS, MIFI},                   
   corref={MISIS},                       
   email={ilia.besedin@gmail.com}   
]{\inits{I}\fnm{Ilya} \snm{Besedin}}
\author[
   addressref={MIFI},
]{\inits{AP}\fnm{Alexey P} \snm{Menushenkov}}


\address[id=MISIS]{
  \orgname{ National University for Science and Technology (MISiS)}, 
  \street{Leninskiy pr. 4},                     %
  \postcode{119049}                                
  \city{Moscow},                              
  \cny{Russia}                                    
}
\address[id=MIFI]{%
  \orgname{National Research Nuclear University MEPhI (Moscow Engineering Physics Institute)},
  \street{Kashirskoe shosse 31},
  \postcode{115409}
  \city{Moscow},
  \cny{Russia}
}



\begin{artnotes}
\end{artnotes}

\end{fmbox}


\begin{abstractbox}

\begin{abstract} 
	We investigate analytically the coupling of a coplanar waveguide resonator to a coplanar waveguide feedline. 
	Using a conformal mapping technique we obtain an expression for the characteristic mode impedances and coupling coefficients of an asymmetric multi-conductor transmission line.
	Leading order terms for the external quality factor and frequency shift are calculated. The obtained analytical results are relevant for designing circuit-QED quantum systems and frequency division multiplexing of superconducting bolometers, detectors and similar microwave-range multi-pixel devices.
\end{abstract}


\begin{keyword}
\kwd{coplanar waveguide}
\kwd{microwave resonator}
\kwd{conformal mapping}
\kwd{coupled transmission lines}
\kwd{superconducting resonator}
\end{keyword}


\end{abstractbox}
%

\end{frontmatter}



\section{Introduction}

Low loss rates provided by superconducting coplanar waveguides (CPW) and CPW resonators are relevant for microwave applications which require quantum-scale noise levels and high sensitivity, such as mutual kinetic inductance detectors \cite{Day2003}, parametric amplifiers \cite{Yamamoto2008}, and qubit devices based on Josephson junctions \cite{Wallraff2004}, electron spins in quantum dots \cite{Childress2004}, and NV-centers \cite{Kubo2011}. 
Transmission line (TL) coupling allows for implementing relatively weak resonator-feedline coupling strengths without significant off-resonant perturbations to the propagating modes in the feedline CPW. Owing to this property and benefiting from their simplicity, notch-port couplers are extensively used in frequency multiplexing schemes \cite{Jerger2012}, where a large number of CPW resonators of different frequencies are coupled to a single feedline. 
The geometric design of such resonators determines the resonant frequencies of their modes, loss rates and coupling coefficients of these modes to other elements of the circuit. 3D electromagnetic simulation software provides excellent means for complete characterization of such structures by finite element analysis. However, in the case of simple structures, analytical formulas can be devised that are invaluable for engineering of large multi-pixel resonator arrays. 

The external quality factor depends not only on the resonator characteristics, but also the coupling elements and port impedances and can be calculated in case of a capacitive \cite{Goppl2008}, inductive \cite{Bothner2013}, and both capacitive and inductive \cite{Khalil2012} couplings. More complex setups with bandpass filters for qubit applications have been developed for capacitive coupling \cite{Jeffrey2014}. In the present article we aim to analytically describe the TL coupled coplanar waveguide (CPW) resonator in an arbitrary-impedance environment. 


Conformal mapping is an established tool in engineering that uses the analytic property of complex-variable functions to transform the boundary shape in boundary value problems of Laplace's equation in two dimensions. Classical results for mode impedances in microstrip and coplanar TL structures can be systematically derived using the Schwarz-Christoffel mapping \cite{Wheeler1964, Wen1969}, yielding explicit formulas in terms of elliptic integrals. The same approach has been successfully applied to finite-width ground plane CPWs and coupled microstrip lines \cite{Veyres1980} and coupled CPWs \cite{Ghione1987}. 
These results rely on the presence of at least one of the structures' symmetry planes, which reduce these problems to two conductors, also yielding explicit formulas with elliptic integrals. 
If the conductor plane symmetry is broken, as in the case of finite-thickness and conductor-backed dielectrics \cite{Ghione1987} and finite-thickness conductors \cite{Veyres1980}, approximate formulas can account for the change in the effective dielectric constant and effective conductor surface area. 

As a rule of thumb, an explicit formula in terms of elliptic integrals for a structure's per-unit-length capacitance and inductance can be obtained, if the boundary value problem for Laplace's equation can be divided into separate decoupled homogeneous domains, and the number of distinct conductors in each domain is no more than two.
In the present paper we use a general approach based on implicit expressions with hyperelliptic integrals \cite{Linner1974,Ghione1994}, recently revisited in \cite{Wang2016}. In contrast to the more established techniques, it is not constrained by the requirement that each domain contains no more than two conductors, but still requires them to be homogeneous. This allows to obtain expressions for the mutual capacitance and inductance matrices of coupled transmission lines even if they have completely arbitrary conductor and gap widths. The effect of finite thickness conductors has been recently investigated \cite{Bertazzi2016}.

\subsection{Resonant scattering}

In the vicinity of a single resonance, the complex scattering parameter as a function of frequency is given by 
\begin{equation}
\label{eq:scattering_single_pole}
S_{kl} \left( f \right)=A_{kl}\left(1+ \frac{B_{kl}}{f-f_{\mathrm{p}}}\right),
\end{equation}
where $A_{kl}$, $B_{kl}$ are arbitrary constants and $f_p$ is the scattering amplitude pole. The resonance frequency is given by the real part $f_p'$ of the pole, while the imaginary part $f_p''$ is proportional to the decay rate.
The total (loaded) quality factor of a resonance is defined as 
\begin{equation}
\label{eq:quality_factor_definition}
Q_l=\left| \frac{f_p'}{2f_p''}\right|.
\end{equation}

For high-quality factor resonances losses are perturbatively small and the total loss rate can be expressed as a sum of loss rates in its components. For a resonator coupled to a TL the losses related to emission of radiation into the TL can be separated from dissipative losses occurring inside the resonator:
\begin{equation}
\frac{1}{Q_l} = \frac{1}{Q_i}+\frac{1}{Q_e}.
\end{equation}
Here $Q_i$ is the internal and $Q_e$ the external quality factors. For most applications, one wishes for the internal quality factor to be as large as possible and for the external quality factor to be as close to the design values as possible. 

From the decay rate of an excitation in the resonant mode one cannot distinguish between loss channels, however external loss can be measured directly by probing the microwave amplitude at the TL ports, or, more practically, connecting the TL to a vector network analyzer and measuring a scattering parameter $S_{21}$.

For a two-port network equation \eqref{eq:scattering_single_pole} can be re-written in terms of real-valued parameters in the form \cite{Khalil2012,Probst2015}
\begin{align}
\label{eq:Sebastian}
S_{21} = 
a e^{i\alpha}e^{2\pi i f \tau}&\left [
1 - \frac{e^{i\phi} Q_l/Q_e}{1+2i Q_l\left(f/f_r-1 \right)}
\right ],
\end{align}
where the additional parameters $a$, $\alpha$, $\tau$ are introduced to characterize transmission through the cables and other components of the measurement system that are not directly connected to the resonator and $Q_l$, $Q_e$ and $f_r$ depend on the coupling element and the resonator. 

The resonance frequency of a CPW resonator is well defined by its length by the formulas
\begin{equation}
\begin{aligned}
\label{eq:fr_0}
f_r^{(0)}=\frac{c_l}{4l}(2p-1), \\
f_r^{(0)}=\frac{c_l}{4l}2p,
\end{aligned}
\end{equation}
where $p=1,2,3,...$ is the mode number and $c_l$ is the speed of light in the medium. 
The first equation corresponds to the case the resonator is terminated with a short-circuit at one end and with an open circuit at the other end ($\lambda/4$ resonator). The second equation corresponds to the case the resonator is terminated either with a short circuit at both ends or with an open circuit at both ends ($\lambda/2$ resonator).

\subsection{Multiconductor transmission lines}  

The TL CPW coupler is a multiconductor TL with finite mutual capacitance and inductance between the resonator conductor and feedline. Wave propagation through a multiconductor TL with $n$ conductors can be described by the multiconductor TL equations \cite{Paul2007}:

\begin{align}
\label{eq:multiconductor-transmission-line}
\frac{\partial \mathbf{V}}{\partial z} = - \mathbf{L}\frac{\partial \mathbf{I}}{\partial t}, \\
\frac{\partial \mathbf{I}}{\partial z} = - \mathbf{C}\frac{\partial \mathbf{V}}{\partial t}.
\end{align}
Here $z$ is the coordinate along the TL, $\mathbf{V}$ and $\mathbf{I}$ the time and coordinate dependent voltages and currents on each of the conductors, and $\mathbf{L}$, $\mathbf{C}$, are the $n \times n$ per-unit-length inductance and capacitance matrices. For simplicity we neglect finite dielectric conductance and conductor resistance as their effect is small for high Q-factor resonators and to first order they only affect the internal quality factor and can be accounted for separately. 

The equations \eqref{eq:multiconductor-transmission-line} can be converted into an algebraic system with a Fourier transform, resulting in an eigenvalue problem
\begin{align}
\label{eq:eigenvalue-multiconductor}
\mathbf{A} \mathbf{b} = \lambda \mathbf{b}
\end{align}
where 
\begin{align}
\label{eq:multiconductor_matrix}
\mathbf{A} = 
\begin{pmatrix}
0 & \mathbf{L} \\
\mathbf{C} & 0
\end{pmatrix}, 
\end{align}
\begin{align}
\mathbf{b} = 
\begin{pmatrix}
\mathbf{V} \\
\mathbf{I}
\end{pmatrix},
\end{align}
and the eigenvalue $\lambda$ is the inverse phase velocity of the propagating mode and the corresponding eigenvector $\mathbf{b}$ is the voltage and current amplitude of the mode. 

For the simplest case of a multiconductor TL system in a homogeneous medium \cite{Paul2007} the per-unit-length inductance and capacitance matrices are up to a constant inverse to each other:
\begin{equation}
\label{eq:homogeneous-media-property}
\mathbf{L}\mathbf{C}=\frac{1}{c_l^2}\mathbf{1},
\end{equation}
where $\mathbf{1}$ is the $n\times n$ identity matrix. If this property holds, $\mathbf{A}$ can be diagonalized with
\begin{equation}
\mathbf{A} = \mathbf{P}\mathbf{D}\mathbf{P}^{-1},
\end{equation}
where 
\begin{align}
\mathbf{P} = 
\begin{pmatrix}
\mathbf{L} & \mathbf{L} \\
c_l^{-1}\mathbf{1} & -c_l^{-1}\mathbf{1}
\end{pmatrix},
\end{align}
and 
\begin{align}
\mathbf{D} = 
\begin{pmatrix}
c_l^{-1}\mathbf{1} & \mathbf{0} \\
\mathbf{0} & -c_l^{-1}\mathbf{1}
\end{pmatrix}.
\end{align}

Another consequence of \eqref{eq:homogeneous-media-property} is that we can define a "characteristic impedance matrix" $\mathbf{Z}$ as 
\begin{equation}
\mathbf{Z} = \mathbf{L}\mathbf{C}^{-1} = \mathbf{C}^{-1}\mathbf{L}.
\end{equation}
$\mathbf{Z}$ is symmetric and can be diagonalized with an orthogonal transform, yielding mode impedances.

\section{Results and discussion}

\subsection{Conformal mapping of the coupled CPW}

Here we follow a general analytical method \cite{Ghione1994} to calculate the per-unit-length capacitance and inductance matrices of two coupled CPWs with arbitrary lateral dimensions and an infinite ground plane. 

We calculate the columns of the capacitance and inverse inductance matrix separately from each other. The geometry of the cross-section of the coupler is shown in fig. \ref{im:cross-section}. In the upper half-plane, the vacuum's relative permittivity and permeability are unity. In the lower half-plane the substrate's dielectric permittivity is $\epsilon$ and the magnetic permeability is $\mu$. On the conductors between the vacuum and the substrate the tangential component of the electric field and normal component of the magnetic field are zero. Due to the problem symmetry, the normal component of the electric field and tangential component of the magnetic field at the vacuum-substrate interface are also zero. The latter property renders the equations for the upper half-plane and the lower half-plane independent from each other. At either of the half-planes, the electric and magnetic fields can be expressed in terms of the electrostatic potential and the component of the magnetostatic vector potential along the TL, which are up to a constant equal to each other.
In either of the half-planes, both the potentials satisfy Laplace's equation:
\begin{equation}
\label{eq:Laplace}
\Delta \phi(x,y) = 0 \
\end{equation}
We set $a_j$ and $b_j$ as the $x$-coordinates of the right and left boundaries of the conductor cross-sections such that $a_0<b_0<a_1<b_1<...<a_n<b_n$, $j=0,...,n$.
If the i-th conductor has a non-zero electric potential $\phi_i$ and all the others are grounded, the boundary conditions at the interface are given by
\begin{equation}
\begin{aligned}
&\phi(x,0) & = & \phi_i, & a_i<x<b_i \\
&\phi(x,0) & = & 0, & a_j<x<b_j, j \neq  i \\
&\frac{\partial \phi}{\partial y}(x,0) & = & 0, & a_j<x<b_{j+1}
\end{aligned}
\end{equation}
In terms of the magnetic scalar potential these boundary conditions correspond to a non-zero magnetic flux $\phi_i$ through any ray originating from the i-th conductor and a zero magnetic flux through any ray originating from any other conductor.
For conformal mapping we introduce the complex variable $z=x+iy$.
Following the general approach, we define the points ${c_j}, j \in \{1,...,n\}\setminus\{i, i+1\}$ such that $a_j<c_j<b_{j}$ where the electric field vector component along the real axis changes its sign. These points must exist since $a_j$ and $b_j$ lie on the $j$-th and $j+1$-th conductors respectively, both of which are grounded, so the integral of the electric field along any contour connecting them must be equal to zero.
The conformal mapping
\begin{equation}
\label{eq:conformal-mapping-Wang}
w(z) =\int\limits_{0}^{z}\frac{\prod\limits_{j \in \{1,...,n\}\setminus{\{i, i+1\}}}{(z-c_j)}} {\prod\limits_{k=0}^{n}{(z-a_k)^{1/2}}\prod\limits_{l=0}^{n}{(z-b_l)^{1/2}}}\mathrm{d}z,
\end{equation}
transforms the conductor surface cross-sections into lines parallel to real axis of the $w$-plane and the vacuum-surface interface into lines parallel to the imaginary axis of the $w$-plane. Due to its analytic property, $\operatorname{Im}{ w(z)}$ satisfies \eqref{eq:Laplace}. The values of $c_j$ are implicitly defined by the boundary conditions at the grounded conductors:
\begin{equation}
\operatorname{Im}w(a_j)=0, j \in \{1,...,n\}\setminus \{i\}.
\end{equation}

The charges induced on each of the conductors induced by the electric potential $\phi_i = \operatorname{Im} {w(a_i)}$ are proportional to their length in the $w$-plane. After accounting for both the substrate and vacuum half-planes and multiplying by the vacuum permittivity (permeability), we obtain for the capacitance matrix and the inverse inductance matrix
\begin{equation}
\begin{aligned}
\label{eq:capacitance_matrix}
C_{ij} = \left(\epsilon+1\right)\epsilon_0\frac{\operatorname{Re}w(b_{j+1})-\operatorname{Re}w(a_j)}{\operatorname{Im}w(a_i)} \\
L^{-1}_{ij} = \left(\frac{1}{\mu\mu_0}+\frac{1}{\mu_0}\right)\frac{\operatorname{Re}w(b_{j+1})-\operatorname{Re}w(a_j)}{\operatorname{Im}w(a_i)} 
\end{aligned}
\end{equation}

Due to the symmetry of this configuration the property \eqref{eq:homogeneous-media-property} holds as if the conductors are inside a homogeneous medium with effective permeability and permittivity: 
\begin{equation}
c_l = \pm\frac{1}{\sqrt{(\epsilon+1)(\mu^{-1}+1)^{-1}\epsilon_0\mu_0}}.
\end{equation}

Finite dimensions of the structure, in particular, finite conductor thickness, have been considered by an additional conformal mapping in a compatible fashion elsewhere \cite{Bertazzi2016}. We do not consider this case here, as it introduces two additional key complications. The first complication is related to the conformal mapping itself, as in this case the boundary value problem cannot be exactly decoupled for the upper and lower half-planes. The second complication is that the relation \eqref{eq:homogeneous-media-property} no longer holds, and different modes of propagation no longer have the same phase velocity. 

In applications the coplanar waveguide coupler is used in either notch-port or butt-port configurations (see fig. \ref{im:tl-schem}). If the coupler length is shorter than the characteristic wavelength of the device one of the modes can be neglected and an effective two-conductor per-unit-length capacitance and inductance matrices can be introduced. The notch-port effective per-unit-length capacitance matrix can be written in terms of the full three-conductor picture as
\begin{align}
\label{eq:notch_port}
\mathbf{C}_\mathrm{n} = 
\begin{pmatrix}
C_{11} & C_{13} \\
C_{31} & C_{33}
\end{pmatrix}.
\end{align}
For the butt-port configuration the effective capacitance matrix is expressed as
\begin{align}
\mathbf{C}_\mathrm{b} = 
\begin{pmatrix}
C_{11}+C_{13}+C_{31}+C_{33} & C_{12}+C_{32} \\
C_{21}+C_{23} & C_{22}
\end{pmatrix}.
\end{align}
The effective two-conductor inductance matrices can be obtained from \eqref{eq:homogeneous-media-property}. Qualitatively, the main effect of finite conductor thickness is an increase to the mutual capacitance of neighboring conductors and corresponding decrease in the inductance. Thus in the notch-port geometry this leads to a decrease in the coupling strength, while in the butt-port to an increase of the coupling strength.

\subsection{Boundary condition equation matrix of the TL model and perturbative analysis}

In order to obtain analytic expressions for the resonance frequencies of a TL system we derive a system of linear equations with unknowns corresponding to all branch currents and nodal voltages and also wave amplitudes in TL elements. The number of modes coexisting at a single frequency inside a multiconductor TL is equal to the dimensionality of \eqref{eq:eigenvalue-multiconductor}, which is twice the number of conductors. 
At each port two boundary conditions link the sum of complex wave amplitudes with nodal voltages and branch currents. 
To link the branch currents together, Kirchhof's branch current law for each node is added to the system.
Finally, ports are added to the system as TLs with a single (outgoing) internal degree of freedom with their characteristic impedance relating the voltage and current amplitudes of this degree of freedom. Incoming waves add to the system of equations as the inhomogeneity. 
The eigenmodes of the system are solutions of the equation
\begin{equation}
\label{eq:transmission-line-model}
\mathbf{M}\mathbf{a} = 0,
\end{equation}
where $\mathbf{M}$ is a square matrix of order $r$, $\mathbf{a}$ is the nodal voltage, branch current and internal degree of freedom amplitude vector.
Poles $f_p$ of the S-matrix as a function of external perturbation frequency $f$ appear whenever the determinant of $\mathbf{M}$ is zero:
\begin{equation}
\label{eq:det_zero}
\Delta(f) = | \mathbf{M} | = 0
\end{equation}
Numerical computation of this determinant is very efficient. However, precise analytical computation of the determinant becomes intractable even for relatively small systems. For a system consisting of independent subsystems, the columns and rows of $\mathbf{M}$ can be permuted such that it becomes a block diagonal matrix with each block corresponding to its own subsystem. In this case the determinant of the entire matrix is the product of the block determinants and can be computed efficiently provided the blocks are small enough.

Coupling introduces non-zero terms in the off-diagonal blocks. Provided that the coupling between the subsystems is weak, the determinant can be expanded in some parameters $\kappa$ characterizing the coupling strength. The $m$-th order of the expansion can be expressed as the sum of of $C(m,r)$ determinants of $\mathbf{M}$ where in all but $m$ of $r$ rows $\kappa$ is zero. Such determinants are more difficult to compute as $m$ blocks merge together, however if $m$ and the individual blocks are small enough computation is still efficient.

The first non-vanishing order $m$ correction of the scattering amplitude poles of the coupled system can be expressed as
\begin{equation}
\label{eq:f_correction}
\Delta f_p = - \sum \limits_\kappa \frac{\frac{\partial^{m} \Delta}{\partial \kappa^{m}}}{\frac{\partial  \Delta}{\partial f}}
\end{equation}
Unlike conventional coupled-mode theory \cite{Barybin2002}, the approach introduced here does not utilize scalar products of mode amplitudes. It can be directly applied to non-Hamiltonian systems where the definition of a scalar product is non-trivial. As a consequence, it can naturally handle cases where coupling is lossy. On the other hand, bare mode coupling coefficients cannot be directly computed in our approach.

The analytical formula for scattering coefficients and nodal voltages and branch currents induced by externally applied signals involves the cofactor matrix to $\mathbf{M}$ and can be computed in a similar fashion. 

\subsection{Boundary condition equation matrix of a TL coupled quarter wavelength resonator}

We consider the TL model shown in fig. \ref{im:tl-schem}. For the coupler we assume that \eqref{eq:homogeneous-media-property} holds. Due to the relatively large number of equations, filling of the equation matrix and all manipulations with it have been carried out by automatic conversion of a SPICE-like description of the TL model in the symbolic algebra system sympy.  To highlight the symmetry between inductance and capacitance and simplify further expressions we express the per-unit capacitance matrix as
\begin{align}
\mathbf{C} = 
\begin{pmatrix}
\gamma c_l^{-1} Z_1^{-1} &-\gamma\kappa  c_l^{-1} (Z_1 Z_2)^{-1/2} \\
-\gamma \kappa  c_l^{-1} (Z_1 Z_2)^{-1/2}&\gamma c_l^{-1} Z_2^{-1},
\end{pmatrix}
\end{align}
and per-unit the inductance matrix as
\begin{align}
\mathbf{L} = 
\begin{pmatrix}
\gamma Z_1 c_l^{-1}&\gamma\kappa (Z_1 Z_2)^{1/2}c_l^{-1} \\
\gamma\kappa (Z_1 Z_2)^{1/2} c_l^{-1}&\gamma Z_2 c_l^{-1} 
\end{pmatrix},
\end{align}
where $\kappa$ is a dimensionless coupling coefficient, $\gamma=\sqrt{1-\kappa^2}$. If $\kappa=0$, $Z_1$ and $Z_2$ become the characteristic impedances of the resonator and the feedline, respectively. The length of the coupler is $l_c$.

The equations relating the port voltages and currents and TL modes inside the coupler are given by
\begin{equation}
\begin{aligned}
\gamma^{-1} c_l V_1 &=& Z_1 A_1 
& + &\kappa \sqrt{Z_1 Z_2} A_2 
& + \\
& & Z_1 A_3 
& + &\kappa \sqrt{Z_1 Z_2} A_4 \\
\gamma^{-1} c_l V_2 &=&\kappa \sqrt{Z_1 Z_2} A_1 
& + & Z_2 A_2
& + \\
& & \kappa \sqrt{Z_1 Z_2} A_3 
& + & Z_2 A_4 \\
\gamma^{-1} c_l V_3 &=& Z_1 \Phi_c A_1 
& + & \kappa \sqrt{Z_1 Z_2} \Phi_c A_2 
& + \\ 
& & Z_1 \Phi_c^{*} A_3 
& + &\kappa \sqrt{Z_1 Z_2} \Phi_c^{*} A_4 \\
\gamma^{-1} c_l V_4 &=& \kappa \sqrt{Z_1 Z_2} \Phi_c A_1 
& + & Z_2 \Phi A_2 
& + \\
& &\kappa \sqrt{Z_1 Z_2} \Phi_c^{*} A_3 
& + & Z_2 \Phi_c^{*} A_4 \\
I_1^{(1)} &=& A_1 & - & A_3\\
I_2^{(1)} &=& \Phi_c A_1 & - & \Phi_c^{*} A_3\\
I_3^{(1)} &=& A_2 & - & A_4\\
I_4^{(1)} &=& \Phi_c A_2 & - &\Phi_c^{*} A_4,
\end{aligned}
\end{equation}
where $\Phi_c=e^{i 2 \pi f l_c/c_l}$ is the phasor that corresponds to the propagation of the microwave from the ports 1 and 2 to the ports 3 and 4.

We set the impedance of the two other resonator sections as $Z_r$ and their lengths as $l_s$ for the shorted end and $l_o$ for the open end. The phase speed in these section is similarly $c_l$. The equations for these section are
\begin{equation}
\begin{aligned}
c_l V_2 & = & Z_r A_5 
& + & Z_r A_6, \\
c_l V_5 & = & Z_r \Phi_s A_5 
& + & Z_r \Phi_s^{*} A_6, \\
I_2^{(2)} &=& A_5 & - & A_6, \\
I_5^{(1)} &=& \Phi_s A_5 & - & \Phi_s^{*} A_6, \\
c_l V_4 & = & Z_r A_7 
& + & Z_r A_8, \\
c_l V_6 & = & Z_r \Phi_o A_7 
& + & Z_r \Phi_o^{*} A_8, \\
I_4^{(2)} &=& A_7 & - & A_8, \\
I_6^{(1)} &=& \Phi_o A_7 & - & \Phi_o^{*} A_8,
\end{aligned}
\end{equation}
where $\Phi_o=e^{i 2 \pi f l_o/c_l}$ and 
$\Phi_c=e^{i 2 \pi f l_s/c_l}$ are phasors corresponding to the propagation of the microwave through the open and shorted ends of the resonator.
Grounding the shorted end yields the equation
\begin{align}
V_5 = 0.
\end{align}
The port impedances at the nodes 1 and 3 are defined as $Z_i$ and $Z_o$, respectively. The corresponding equations are
\begin{equation}
\begin{aligned}
V_1 &= -Z_i A_9, \\
I_1^{(2)} &= -A_9, \\
V_3 &= -Z_i A_{10}, \\
I_3^{(2)} &= -A_{10}.
\end{aligned}
\end{equation}
Finally, the sum of currents flowing into each node should equal to zero. This gives 
\begin{equation}
\begin{aligned}
I_i^{1}+I_i^{2} = 0, i=1,...,5, \\
I_6 = 0
\end{aligned}
\end{equation}
The unknowns in these equations are
\begin{equation}
\begin{aligned}
I_i^{(j)}, i=1,...,5, j=1,2; \\
I_6, \\
V_k, k=1,...,6, \\
A_l, l=1,...,10.
\end{aligned}
\end{equation}
With 27 unknowns and 27 equations, the system matrix is square. 

\subsection{Quality factor and frequency shift of a transmission-line-coupled resonator}

We consider the general cases of a CPW resonator coupled to a TL (fig. \ref{im:tl-schem}, d) with termination impedances (a) $Z_{t1}=0$, $Z_{t2}=\infty$, (b) $Z_{t1}=0$, $Z_{t2}=0$ and (c) $Z_{t1}=\infty$, $Z_{t2}=\infty$. Both the notch-port and capacitive and inductive butt-port coupled CPW resonators are described with this schematic. 
 
Following the approach introduced in the previous section, we expand the determinant $\Delta$ to first order in $Z_2-Z_r$ and to second order in $\kappa$. 

To the zeroth order, the solution of $\Delta = 0$ yields the poles of an isolated resonator \eqref{eq:fr_0} and the poles corresponding to standing waves in the feedline that arise due to impedance mismatch between the feedline and the input and output ports. The equation for these poles is
\begin{equation}
Z_{f} \left( Z_{i} + Z_{0} \right) \cos{\theta} -i \left( Z_{f}^{2} + Z_{i} Z_{o} \right) \sin{\theta} = 0.
\end{equation}
The derivatives of the determinant of the system (a) with respect to $f$, $\kappa$ and $Z_2$ evaluated at the the $p$-th resonance frequency are given by

\begin{equation}
\label{eq:dD/df^a}
\begin{aligned} 
\frac{\partial \Delta^a}{\partial f} = & \frac{32 \pi}{c_{l}} \left(-1\right)^{p} Z_{r}^{3} \left(l_{c} + l_{o} + l_{s}\right) \times \\
\Big(
& - Z_{f} \left( Z_{i} + Z_{0} \right) \cos{\theta} +i \left( Z_{f}^{2} + Z_{i} Z_{o} \right) \sin{\theta} \Big), \\
\frac{\partial \Delta^a}{\partial Z_2} = &- 16 \left(-1\right)^{p} Z_{r}^{2}  \sin{\theta} \cos{\psi}   \times \\
\Big( & - Z_{f} \left(Z_{i} + Z_{o}\right) \cos{\theta} + i \left(Z_{f}^{2} + Z_{i} Z_{o}\right) \sin{\theta} \Big), \\
\frac{\partial \Delta^a}{\partial \kappa} = & 0, \\
\frac{\partial^2 \Delta^a}{\partial \kappa^2} = &16 \left(-1\right)^{p} Z_{r}^{3} \sin{\theta}   \times \\
\Big(
&Z_{f} \left( Z_{o} - Z_{i}\right) \sin{\psi} \sin{\theta} - \\
&Z_{f} \left(Z_{i} + Z_{o}\right) \left(2 \cos{\psi} \cos{\theta} + 1\right) + \\
& i \left(3 Z_{f}^{2} + Z_{i} Z_{o}\right) \sin{\theta} \cos{\psi}
\Big),
\end{aligned}
\end{equation}
where $\theta$ and $\psi$ are phase variables that can be calculated from the zeroth-order resonance frequency \eqref{eq:fr_0} with
\begin{equation}
\begin{aligned}
\label{eq:psi_theta}
\psi = & 2\pi \left(l_c + 2l_o\right)f^{(0)}_r/c_l, \\
\theta = & 2\pi l_c f^{(0)}_r/c_l.
\end{aligned}
\end{equation}
The frequency shift and decay rate of the resonator due to the presence of the TL can be obtained by substituting these values into \eqref{eq:f_correction}. For a matched input port $Z_i=Z_f$ and matched output port $Z_o=Z_f$ we obtain 
\begin{equation}
\begin{aligned}
\label{eq:mm_l4}
\Delta f^{a}_r = & - \frac{c_{l} \kappa^{2} \sin{\theta}\left(2  \cos{\psi} +  \cos{\theta} \right)}{4 \pi \left(l_{c} + l_{o} + l_{s}\right)} \\ -
& \frac{c_{l}\left(Z_{2}-Z_{r}\right)\sin{\theta}\cos{\psi}}{2 \pi Z_{r} \left(l_{c} + l_{o} + l_{s}\right)}, \\
\frac{1}{Q^{a}} = & \frac{2 \kappa^{2} \sin^{2}{\theta}}{\pi \left(2 p - 1\right)}.
\end{aligned}
\end{equation}

Similarly, for the resonator terminations (b) with shorted terminations on both ends the derivatives of the determinant are given as

\begin{equation}
\label{eq:dD/df^b}
\begin{aligned}
\frac{\partial \Delta^{b}}{\partial f} = & - \frac{32 \pi i}{c_{l}} \left(-1\right)^{p} Z_{r}^{4} \left(l_{c} + l_{o} + l_{s}\right) \times \\ & \Big( Z_{f} \left(Z_{i}+Z_{o}\right) \cos{\theta } -i \left(Z_{f}^{2}+ Z_{i} Z_{o}\right) \sin{\theta} \Big), \\
\frac{\partial \Delta^b}{\partial Z_2} = & - 16 i \left(-1\right)^{p} Z_{r}^{3} \sin{\left (\theta \right )} \cos{\left (\psi \right )} \times \\
& \Big(Z_{f} \left(Z_{i} + Z_{o}\right) \cos{\theta} - i \left(Z_{f}^{2} + Z_{i} Z_{o}\right) \sin{\theta}\Big), \\
\frac{\partial \Delta^b}{\partial \kappa} = & 0, \\
\frac{\partial^2 \Delta^b}{\partial \kappa^2} = & 16 i \left(-1\right)^{p} Z_{r}^{4}  \sin{\theta}\times \\
\Big(
& Z_{f} \left(Z_{o} - Z_{i}\right) \sin{\psi} \sin{\theta} - \\
& Z_{f} \left(Z_{o} + Z_{i}\right) \left(2 \cos{\psi} \cos{\theta} - 1\right) + \\ 
& i\left(3 Z_{f}^{2} + Z_{i} Z_{o}\right) \sin{\theta} \cos{\psi}
\Big),
\end{aligned}
\end{equation}
For a matched input port $Z_i=Z_f$ and matched output port $Z_o=Z_f$ we obtain 
\begin{equation}
\begin{aligned}
\Delta f^{b}_r = & \frac{c_{l} \kappa^{2} \sin{\theta} \left(2 \cos{\psi} - \cos{\theta}\right) }{4 \pi \left(l_{c} + l_{o} + l_{s}\right)} \\ - 
& \frac{c_{l} \left(Z_{2} - Z_{r}\right) \sin{\theta} \cos{\psi} }{2 \pi Z_{r} \left(l_{c} + l_{o} + l_{s}\right)}, \\
\frac{1}{Q^{b}} = & \frac{\kappa^{2} \sin^{2}{\theta }}{\pi p}.
\end{aligned}
\end{equation}

Finally, for the terminations (c) we have

\begin{equation}
\label{eq:dD/df^c}
\begin{aligned}
\frac{\partial \Delta^{c}}{\partial f} = & - \frac{32 \pi i}{c_{l}} \left(-1\right)^{p} Z_{r}^{2} \left(l_{c} + l_{o} + l_{s}\right) \times \\
& \Big(Z_{f} (Z_{o}+Z_{i}) \cos{\theta} - i \left(Z_{f}^{2} + Z_{i} Z_{o}\right) \sin{\theta} \Big), \\
\frac{\partial \Delta^c}{\partial Z_2} = & 16 \left(-1\right)^{p} i Z_{r} \sin{\left (\theta \right )} \cos{\left (\psi \right )} \times \\
\Big( & Z_{f} \left(Z_{i} + Z_{o}\right) \cos{\theta} - i \left(Z_{f}^{2} + Z_{i} Z_{o}\right) \sin{\theta}\Big), \\
\frac{\partial \Delta^c}{\partial \kappa} = & 0, \\
\frac{\partial^2 \Delta^c}{\partial \kappa^2} = & - 16 i \left(-1\right)^{p} Z_{r}^{2} \sin{\theta} \times \\ 
\Big(
& Z_{f} \left(Z_{o} - Z_{i}\right) \sin{\psi} \sin{\theta} + \\
& Z_{f} \left(Z_{o} + Z_{i}\right) \left(2 \cos{\psi} \cos{\theta} + 1\right) - \\
& i \left(3 Z_{f}^{2} + Z_{i} Z_{o}\right) \sin{\theta} \cos{\psi}
\Big)
\end{aligned}
\end{equation}
For a matched input port $Z_i=Z_f$ and matched output port $Z_o=Z_f$ 
\begin{equation}
\begin{aligned}
\Delta f^{c}_r = & - \frac{c_{l} \kappa^{2} \sin{\theta}\left(2 \cos{\psi} + \cos{\theta}\right)}{4 \pi \left(l_{c} + l_{o} + l_{s}\right)} \\ & - \frac{c_{l} \left(Z_{2} - Z_{1}\right) \sin{\theta}\cos{\psi}}{2 \pi Z_{r} \left(l_{c} + l_{o} + l_{s}\right)}, \\
\frac{1}{Q^{c}} = & \frac{\kappa^{2} \sin^{2}{\theta }}{\pi p}.
\end{aligned}
\end{equation}

In the case of matched input and output port impedances the quality factor has no leading-order dependence on the position of the coupler section. This result arises due to the equal contribution of inductive and capacitive coupling in \eqref{eq:homogeneous-media-property} and equal current and voltage amplitude in the feedline. For unmatched ports, this symmetry is broken and standing waves in the coupler section of the feedline can both increase and decrease the quality factor of the resonator. 

\subsection{Comparison with numerical simulation}
To verify and justify the theoretical result obtained above, we have performed a finite-element 3D electromagnetic simulation of the scattering parameter of a circuit containing a $\lambda/4$ CPW resonator on a silicon substrate shown in fig. \ref{im:3D} as a function of the width of the electrode between the resonator and the feedline $w_3$ with the commercial solver ``HFSS''.
For the simulation we set numerical values of the material properties $\frac{\epsilon+1}{2}=6.225$ and $\mu=1$. The conductor and gap width of the feedline is $w_1=16 ~\mathrm{\mu m}$ and $s_1=8 ~\mathrm{\mu m}$ and the input and output ports are impedance matched ($Z_i=Z_o=48.33 ~\Omega$). The conductor and gap width of the resonator was $w_2=7 ~\mathrm{\mu m}$ and $s_2=4 ~\mathrm{\mu m}$, which corresponds to $Z_r=50.22 ~\Omega$. The length of the coupler section is $l_c=400 ~\mathrm{\mu m}$, the length of the shorted section of the resonator CPW is $l_s=3600 ~\mathrm{\mu m}$ and of the open end $l_o=1000 ~\mathrm{\mu m}$. 
To obtain the quality factor and resonance frequency of the resonator, we have fitted the simulated scattering parameter as a function of frequency with the formula \eqref{eq:Sebastian} according to the procedure described in \cite{Probst2015}. 

For the TL model calculation the inductance and capacitance matrices of the coupler section have been extracted using the formulas \eqref{eq:capacitance_matrix} and \eqref{eq:notch_port}. The comparison of the 3D simulation results, the numerical solution of the equation \eqref{eq:det_zero}, and the approximate formulas \eqref{eq:mm_l4} for the fundamental mode of the resonance is shown in fig. \ref{im:comparison}. The analytical first-order approximation to the TL model shows little deviation from the precise numerical result for the TL.

Compared to the 3D simulation, the TL model yields systematically higher estimates of the quality factor.  This can be attributed to the presence of spurious coupling between the resonator and feedline, primarily of the resonator conductor arcs attached to the coupler, which leads to a larger effective coupler length $l_c$. The error of the quality factor calculation is within 70\% over the entire range of simulated quality factors, which spans over three orders of magnitude from $10^3$ to $10^6$. The deviation can be reduced by increasing the length of the coupler section while reducing the conductor width $w_3$ and reducing the arc radius. The small frequency shift dependence on $w_3$ predicted by our analytical model cannot be reliably reproduced with our 3D simulation due to precision and meshing issues even for large simulation sizes.

Apart from the errors that can be identified and quantified by EM simulation, practical devices are also plagued by the infamous problem of standing waves. Non-ideal connectors lead to frequency-dependent port impedances, which in turn enter the formulas \eqref{eq:dD/df^a}, \eqref{eq:dD/df^b}, \eqref{eq:dD/df^c} and lead to unpredictable variations of the quality factor. This issue, if not properly handled by input and output RF circulators or isolators, or use of high-precision connectors, is arguably the main source of quality factor deviations in devices.

\section{Conclusions}
In this article we have presented an analytical model for transmission line coupled high quality factor coplanar waveguide resonators. We consider two main configurations, the butt-port and notch-port coupled CPW resonators. Using a perturbative expansion of the equation for the resonance frequencies, we obtain the external quality factors and frequency shifts for $\lambda/2$ and $\lambda/4$ resonators due to coupling to an arbitrary-impedance environment. 3D simulation of a sample design shows that the method's accuracy is limited by spurious couplings in the actual layout. 

The obtained analytical results can be applied to accelerate the design of large circuit-QED quantum systems and frequency division multiplexing of superconducting bolometers, detectors and similar microwave-range multi-pixel devices, eliminating the need for full 3D electrodynamic simulations.


\begin{backmatter}

\section*{Acknowledgments}
The authors would like to thank A. Ustinov for sharing his comments on an early version of the manuscript, and V. Chichkov, G. Fedorov, I. Khrapach for their efforts on the design, fabrication and characterization of CPW resonators, which have motivated this work.
\section*{Availability of data and material}
Example calculations and 3D EM simulation results are available at  \url{https://github.com/ooovector/cpw_coupling}. Source code for reproducing the main analytical results is included in the supplementary.
\section*{Competing interests}
The authors declare that they have no competing interests.
  
\section*{Funding}
This work was supported by the Russian Science Foundation (grant No. 16-12-00095).

\section*{Authors' contributions}
The authors' contributions are equal.

\bibliographystyle{bmc-mathphys} 
\bibliography{coupling_paper}      




\section*{Figures}
  \begin{figure}[h!]
  		\includegraphics{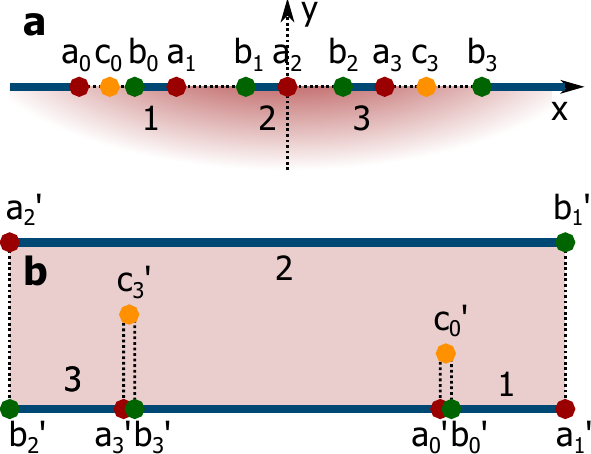}
  		\caption{Conformal mapping of the coupler cross-section.}
  		(a) Original $z$-plane. All of the points $a_j$, $b_j$, and $c_j$ are real and located on the real axis. (b) Transformed $w$-plane. The points $a'_j$, $b'_j$ and $c'_j$ are the transforms of $a_j$, $b_j$, and $c_j$, respectively. In the notch-port configuration, conductor 1 is the feedline, conductor 2 is grounded and conductor 3 is the resonator, and both the feedline and resonator can extend from both ends of the coupler. In the butt-port configuration, conductors 1 and 3 are terminated on one end of the coupler. On the other end conductor 2 is terminated, and conductors 1 and 3 are connected to each other. 
  		\label{im:cross-section}
  	\end{figure}

\begin{figure}[h!]
  	\includegraphics[width=0.5\paperwidth]{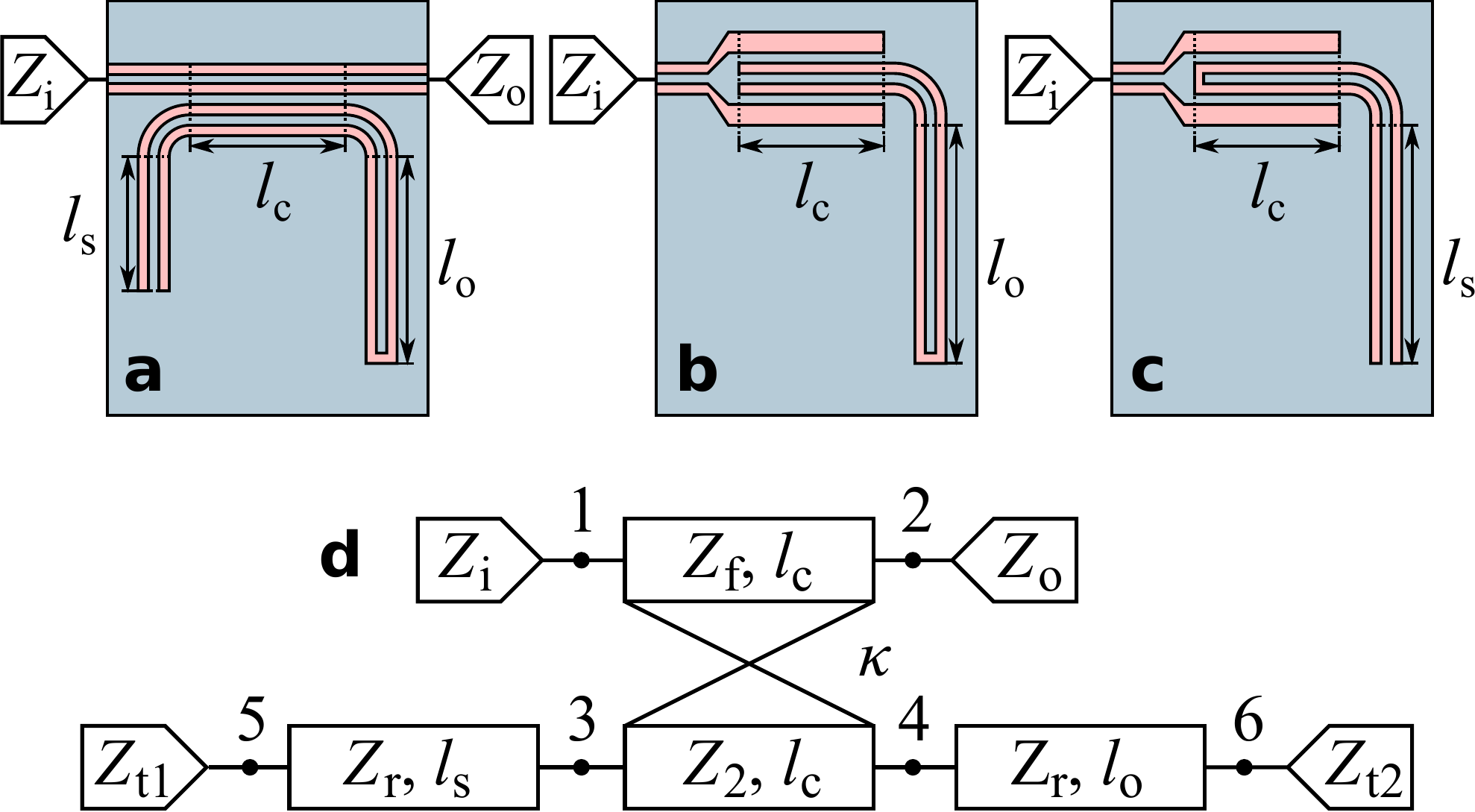}
  \caption{Schematic of a notch-port and butt-port coupled resonators.}
Top-down view. Metalized areas of the structure are shown in grey, non-metalized are shown in pink. The feedline is connected to transmission lines with impedances $Z_\mathrm{i}$ and $Z_\mathrm{o}$ (not to scale). The length of the coupler, open and shorted sections are $l_\mathrm{c}$, $l_\mathrm{o}$, $l_\mathrm{s}$, respectively.
(a) $\lambda/4$ CPW resonator notch-port coupled to a transmission line, (b) $\lambda/4$ CPW resonator inductively butt-port coupled to a transmission line, (c) $\lambda/4$ CPW resonator capacitively butt-port coupled to a transmission line with input impedance $Z_{\mathrm{i}}$. (d) TL schematic of the general case with arbitrary terminations of the resonator $Z_{\mathrm{t1}}$ and $Z_{\mathrm{t2}}$ and feedline $Z_{\mathrm{o}}$ and $Z_{\mathrm{i}}$.
\label{im:tl-schem}
\end{figure}

\begin{figure}[h!]
	\includegraphics[width=0.5\paperwidth]{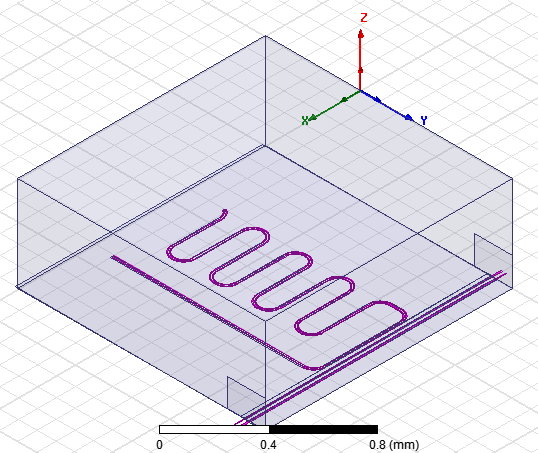}
\caption{CPW resonator and feedline model used for 3D numerical simulation.} 
	The isolated sections of the resonator waveguide are bent into a meander for compactness. Only the length of the straight coupler section is included in $l_c$. The lengths of the arcs are included into the lengths of the shorted and open ends of the resonator $l_s$ and $l_o$.
\label{im:3D}
\end{figure}

\begin{figure}[h!]
	\includegraphics{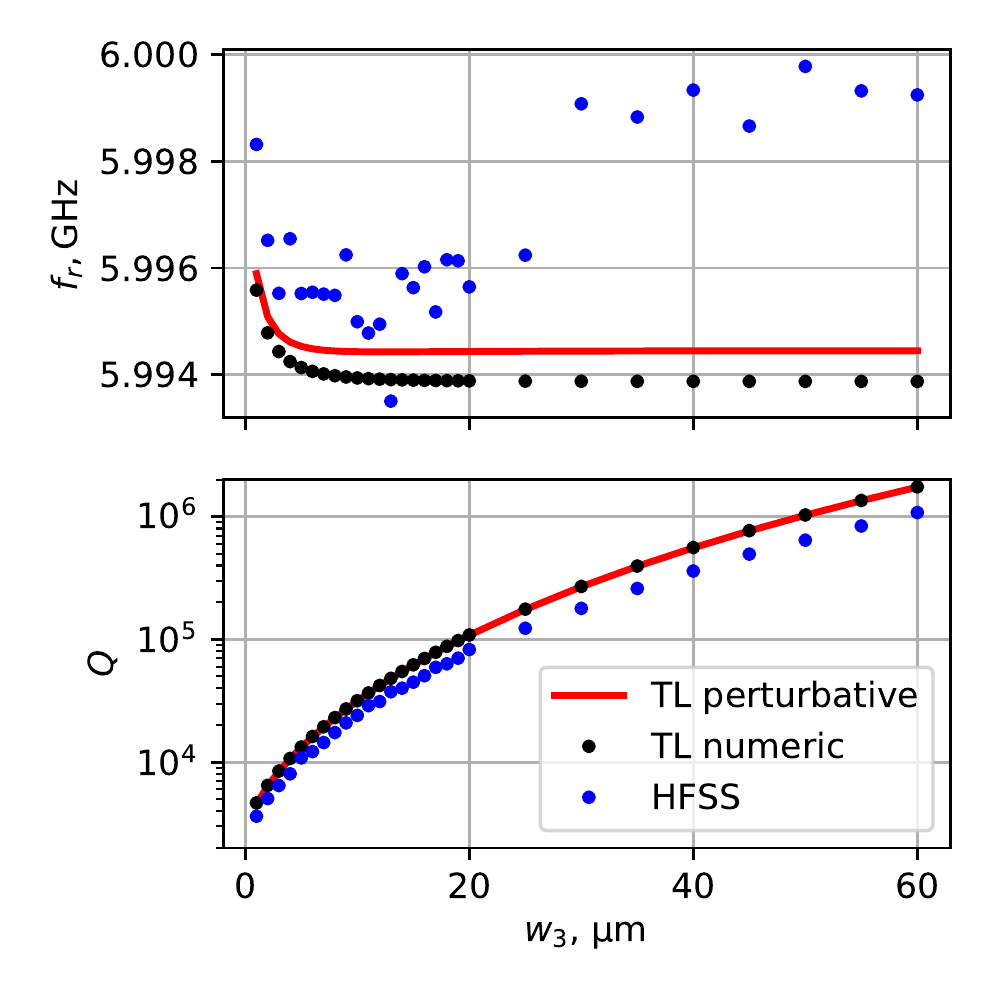}
\caption{Comparison with full 3D electromagnetic simulation.} Quality factor and resonance frequency of the fundamental mode calculated by full 3D simulation (HFSS), using formula \ref{eq:mm_l4} (TL formula), and by numeric solution of \eqref{eq:det_zero}.
\label{im:comparison}
\end{figure}

\end{backmatter}
\end{document}